%% file: main.tex
\documentclass{egpubl}
\usepackage{eurovis2023}

\SpecialIssuePaper         

\CGFStandardLicense

\usepackage[T1]{fontenc}
\usepackage{dfadobe}  

\usepackage{cite}  
\BibtexOrBiblatex
\electronicVersion
\PrintedOrElectronic
\ifpdf \usepackage[pdftex]{graphicx} \pdfcompresslevel=9
\else \usepackage[dvips]{graphicx} \fi

\usepackage{egweblnk}

\usepackage{eurovis2023}
\usepackage{xcolor}
\usepackage{tikz}
\usepackage{multirow}
\usepackage{ulem}
\usepackage{egweblnk}

\newcommand\alexout{\bgroup\markoverwith{\textcolor{red}{\rule[0.5ex]{2pt}{0.4pt}}}\ULon}

\title{Don't Peek at My Chart: Privacy-preserving Visualization for Mobile Devices}

\author[Songheng Zhang , Dong Ma \& Yong Wang]
{\parbox{\textwidth}{\centering Songheng Zhang$^{1}$\orcid{0000-0002-0191-220X}
        ,  Dong Ma$^{1}$\orcid{0000-0003-3824-234X} and Yong Wang$^{1}$\thanks{Y. Wang is the corresponding author.}\orcid{0000-0002-0092-0793}
        }
        \\
{\parbox{\textwidth}{\centering $^1$Singapore Management University, Singapore\\
       }
}
}
        
\begin{document}
\maketitle
\begin{abstract}

Data visualizations have been widely used on mobile devices like smartphones for various tasks (e.g., visualizing personal health and financial data), 
making it convenient for people to view
such data anytime and anywhere.
However, others nearby can also easily peek at the visualizations, resulting in personal data disclosure. In this paper, we propose a perception-driven approach to transform mobile data visualizations into privacy-preserving ones.
Specifically, based on human visual perception, we
develop a masking scheme to adjust the spatial frequency and luminance contrast of colored visualizations. 
The resulting visualization retains its original 
information in close proximity but reduces visibility when viewed from a certain distance or farther away. We conducted two user studies to inform the design of our approach (N=16) and systematically evaluate its performance (N=18), respectively. The results demonstrate the effectiveness of our approach in terms of privacy preservation for mobile data visualizations.

\end{abstract}

\input{sources/1-introduction.tex}
\input{sources/2-related_work}

\input{sources/3-background.tex}
\input{sources/4-method.tex}
\input{sources/5-experiment.tex}
\input{sources/6-discussion.tex}
\input{sources/7-conclusion.tex}

\section*{Acknowledgment}
This research was supported by Lee Kong Chian Fellowship and the Singapore Ministry of Education (MOE) Academic Research Fund (AcRF) Tier 2 grant (Grant number: T2EP20222-0049).
We would like to thank the participants in our user study and the anonymous reviewers for their valuable feedback.

\bibliographystyle{eg-alpha-doi}  
\bibliography{egbibsample}

\input{sources/8-appendix}

\end{document}

%% file: sources/1-introduction.tex
\section{Introduction}

\begin{figure}[htbp!]
    \centering
    \includegraphics[width=0.8\linewidth]{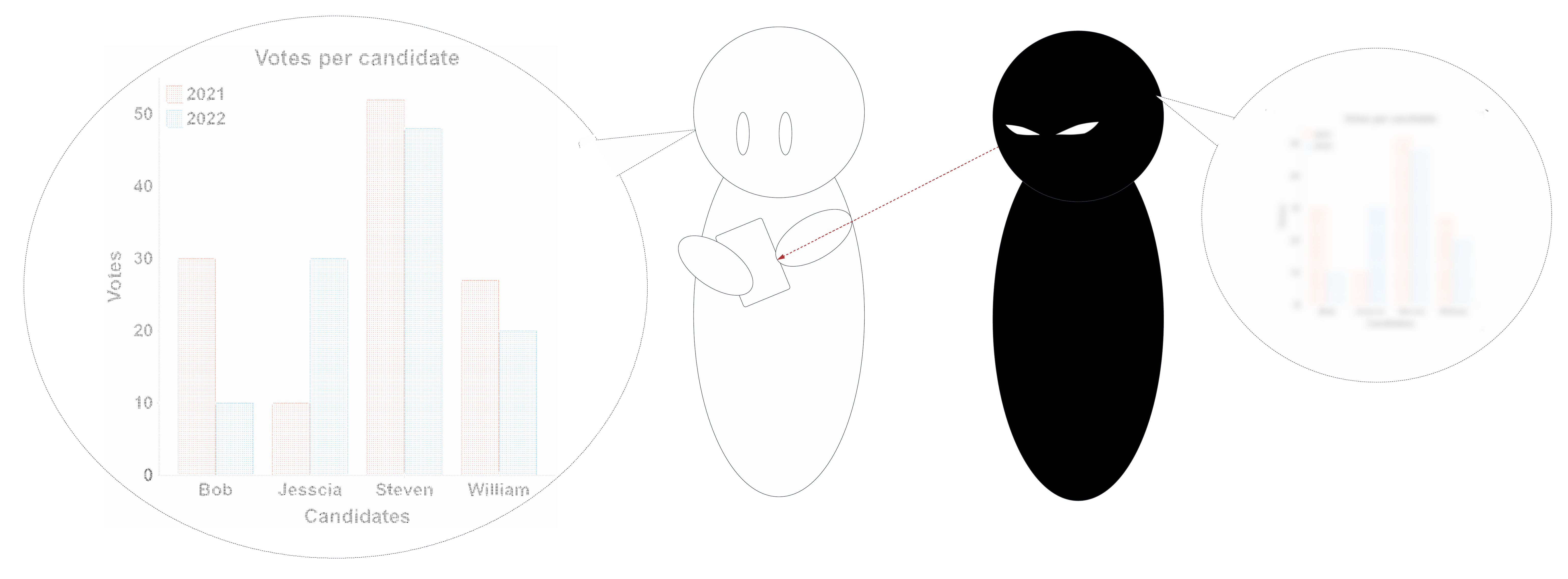}
    \caption{An example application scenario of our method. The data owner (left side) in proximity can see the privacy-preserving visualization displayed on the smartphone, while the shoulder surfer (right side) at a far distance can not interpret the visualization content.}
    \vspace{-2em}
    \label{fig:image_sp}
\end{figure}

Data visualization is ubiquitous for data exploration and analysis.
With the explosive popularity of mobile devices such as smartphones and smartwatches,
they have become one of the major platforms for people to view visualizations of various data~\cite{lee2021mobile}, which are often sensitive personal data.
For example, people may check a bar chart on a healthcare app showing one's health and fitness data, like sleeping hours and walking steps~\cite{applewatch, garmin}, and explore a pie chart via one's mobile bank service displaying their portfolio components~\cite{BoA}.
Unlike traditional desktop-based visualizations often used in a private environment (e.g., in the office or at home), mobile data visualizations can be explored by users anytime and anywhere.
Accordingly,
data privacy issues arise when people are viewing mobile data visualizations showing sensitive personal data in public areas (e.g., on a bus or a train), as shoulder surfers nearby can also easily peek at those visualizations displayed on the mobile device screen, which is commonly known as \textit{Shoulder Surfing} as shown in Figure~\ref{fig:image_sp}. Additionally, \textit{shoulder surfers} are individuals who engage in shoulder surfing~\cite{abdrabou2022understanding}.
Though it is possible to be more cautious when using mobile data visualizations in public space, prior research has shown that only $7\%$ of people are aware of  Shoulder Surfing cases in the wild~\cite{eiband2017understanding}.

The most popular way to preserve data privacy on mobile devices is to attach a privacy film to their screens.
By restricting the visible range of a screen to a particular viewing angle, it can effectively prevent personal information leakage when shoulder surfers nearby are not within the visible range of angles.
Nevertheless, it takes extra cost to purchase a privacy film to attach it to the device screen.
Also, privacy can negatively affect the sensitivity of the screen and lower the visibility of all the applications on mobile devices~\cite{ali2014protecting}, whether the corresponding application data is sensitive or not.
Prior visualization research has started to address the data privacy issues of visualization~\cite{bhattacharjee2020privacy}, but mostly focuses on anonymizing the underlying individual data items. It is achieved by introducing uncertainty to either the data space~\cite{chou2016privacy,chou2019privacy,wang2017utility} or the visual mapping between data items and visual encodings~\cite{dasgupta2012conceptualizing,chou2016obfuscated,dasgupta2019guess,wang2015ambiguityvis}, making individual items indistinguishable and preventing identity and attribute disclosures.
However, these approaches cannot achieve privacy preservation for
the visualization itself, i.e., hiding the visualization from Shoulder Surfing,
which is also crucial for data privacy preservation.
Two recent promising studies, IllustionPin~\cite{papadopoulos2017illusionpin} and HideScreen~\cite{chen2019keep}, attempt to protect information on mobile devices.
They either utilize the concept of \textit{hybrid image}~\cite{papadopoulos2017illusionpin} to hide the PIN password~\cite{chen2019keep} or discretize the device screen into grid patterns to make on-screen texts or personal grayscale images blend into the black background.
However, they target grayscale natural images or PIN keypad images and cannot work for data visualizations that are often colorful~\cite{silva2011using} and involve different visual marks~\cite{munzner2014visualization}, like rectangles, circles, and lines.



To fill the research gap,
we propose a privacy-preserving visualization approach for mobile devices, which can guarantee the visibility of the visualization at proximity but hide visualizations when viewed from a certain distance or above (Figure~\ref{fig:image_sp}).
Our approach is inspired by the study of human visual perception~\cite{barten1999contrast}, which suggests that spatial frequency and luminance contrast primarily determine human visibility of visual stimuli~\cite{national1985emergent}. We also take into account the different properties of various visual marks~\cite{munzner2014visualization} and are informed by our preliminary user study to develop customized masking schemes for line-based marks (e.g., line, text, and axes) and area-based marks (e.g., circles and rectangles) in visualizations to enhance the privacy preservation while maintaining proximity visibility. Specifically, our method is a masking scheme that consists of two granularity  levels of operations aimed at enhancing  the privacy preservation of mobile data visualization. At the coarse-grained level, the method transforms the spatial frequency and luminance contrast of the input visualization image. This operation adjusts the visualization visibility and ensures shoulder surfers cannot access the visualization at a distance. Further, at the fine-grained level, we consider different characteristics of visual marks
and propose customized masking schemes for them, balancing the visualization visibility at a close proximity and privacy preservation at a distance.


There is a large design space to alter masking schemes and luminance contrast for privacy preservation of mobile data visualizations.
To narrow the design scope, we first conducted a preliminary user study with 16 participants to identify suitable variable design choices. The study reveals that participants' perception of the resulting visualizations depends on the mask area or the luminance contrast between the visual marks and background.
A suitable range of mask areas and luminance contrast can achieve the best trade-off between visualization visibility in close proximity (e.g., 30cm) and privacy preservation at a far distance (e.g., 60cm).
Also, the participants' feedback has motivated us to design adaptive fine-grained masking schemes for different visual marks.


Guided by our preliminary study's findings,
we further carried out a user study with another set of 18 participants to evaluate the complete version of our approach with the suitable variable configurations in comparison with baseline approaches (i.e., the original visualizations and the partial version of our approach).
The results show that our approach can achieve similar visibility with the original visualizations in  proximity, but 
enhance the privacy preservation of data visualizations when being viewed from a certain distance (i.e., 90cm), demonstrating its usefulness and effectiveness in terms of visualization privacy preservation on mobile devices.

In summary, the main contributions of this paper can be summarized as follows:
\begin{itemize}
    \item We propose a novel perception-driven approach
    for privacy-preserving mobile data visualization
    by adjusting the spatial frequency and luminance contrast of visual marks, which, to the best of our knowledge, is the first of its kind.

    \item We conducted two
    user studies to inform the design of our approach and further extensively evaluate our approach in comparison with baseline approaches across different types of charts. The results demonstrate our approach's usefulness and effectiveness in terms of privacy preservation.

  \item We summarize the lessons we learned during the development of the proposed approach, which
  can shed light on future research on the privacy-preservation of mobile data visualizations.
\end{itemize}

%% file: sources/2-related_work.tex
\section{Related work}




\subsection{Data Anonymization in Visualization}

Protecting sensitive information from potential leakage is a critical task for many real-world tasks
~\cite{bhattacharjee2020privacy}.
With the wide application of data visualization, there has been a growing trend of research on the privacy preservation of data visualizations.
Specifically, many
data anonymization techniques have been developed to encrypt identifiers of individual data items that can connect individual data items to
different visual elements
in a visualization~\cite{oksanen2015methods,wang2018graphprotector,dasgupta2012conceptualizing,dasgupta2019guess}.
According to Bhattacharjee~\textit{et al.}~\cite{bhattacharjee2020privacy}, these data anonymization techniques can be divided into two groups: (1) data uncertainty and (2) visual uncertainty. Data uncertainty-based methods remove or modify a portion of the original dataset to ensure that a specific number of data records are indistinguishable and thereby stop sensitive exposure of sensitive individual data items~\cite{sweeney2002k,bayardo2005data}.

For example, Chou \textit{et al.}~\cite{chou2019privacy} proposed a data-based clustering algorithm on sequential data and prevented the individual data items from being revealed by visualization (e.g., the Sankey diagram).
Similarly, Okansen \textit{et al.}~\cite{oksanen2015methods} introduced a privacy-preserving heatmap for trajectory data, where three data processing techniques are presented to prevent the data owner's identity disclosure from being identified in the heatmap. 
In contrast to data uncertainty, visual uncertainty-based data anonymization approaches involve uncertainty in the mapping between data points and visualizations.
For instance, Dasgupta~\textit{et al.}~\cite{dasgupta2013measuring} proposed a method for clustering similar data points, binning them according to their value ranges, and displaying these data in parallel coordinate views. Chou~\textit{et al.}~\cite{chou2016obfuscated} obfuscated data rendering in scientific visualizations, and the visualizations can prevent unauthorized viewers from seeing the detailed data items.
Although these methods can avoid the revelation of individual data items,
they cannot prevent shoulder surfers from viewing the visualization itself.


\subsection{Shoulder-surfing Protection on Mobile Devices}


According to Chen~\textit{et al.}~\cite{chen2019keep},
prior studies on shoulder-surfing protection can be categorized into three types: interpretation barrier, shoulder surfer alter, and information blocking.
The interpretation barrier method does not stop shoulder surfers from observing sensitive information but slows down the information leakage on the mobile screen. As a result, sensitive data is displayed in a specific format that the data owner can only understand. For example, Von Zezschwitz~\textit{et al.}~\cite{von2016you} recommended using graphic distortion filters to protect private images in smartphone photo galleries. Distortion filters made it easy for the user (i.e., the image owner) to recognize the content but difficult for the shoulder surfer to comprehend. Similarly, Gouveia~\textit{et al.}~\cite{gouveia2016exploring} designed an alternative metaphor (i.e., a growth garden) to represent physical activity progress. Alternatively, the methods of shoulder surfer alter utilize sensors in a mobile device to detect if anyone is nearby and try to peek at the user's mobile device screen~\cite{ryu2017electronic,ali2014protecting}. For example, Ryu~\textit{et al.}~\cite{ryu2017electronic} leverages face recognition technology and the front camera to detect whether people around the user are peeking at the mobile screen.
The information blocking method aims to conceal the actual information on the screen when shoulder surfers view it. For instance, adding a privacy film to a mobile device screen can protect users' information on the screen, but it incurs additional expenses.  
In contrast to the hardware method, IllusionPIN~\cite{papadopoulos2017illusionpin} blends real and fake keypads using the hybrid image~\cite{oliva2006hybrid}. As a result, users can read the actual keypad, while shoulder surfers can only see the fake keypad.  In addition to PIN keypad protection, HideScreen utilizes the optical system to blend screen content into the background, thus preventing shoulder surfing~\cite{chen2019keep}. Nevertheless, IllusionPIN is only available for a specific image type (i.e., mobile keypads), and HideScreen only works with grayscale nature images that sacrifice color information.
Unlike prior studies above, our approach targets privacy preservation for mobile data visualizations, which are colorful and different from natural images.

%% file: sources/3-background.tex
\section{Background}
\label{sec:background}
In this section, we introduce the background of our research, including image frequency (Section~\ref{background:sp}), luminance contrast (Section~\ref{background:contrast}), and contrast sensitivity function (Section~\ref{background:csf}).

\begin{figure}[ht!]
    \centering
    \includegraphics[width=0.8\linewidth]{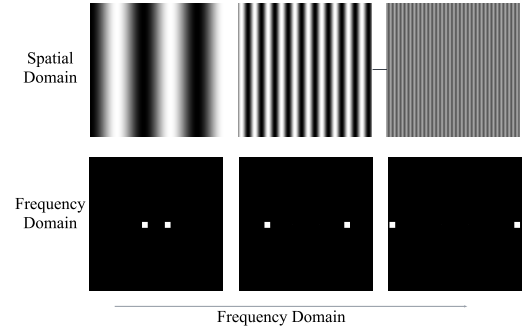}
    \caption{Representations of images (gratings) in the spatial  domain (upper row) which is composed of white and black bars, and frequency domain (lower row) in which white squares represent the spatial frequencies. The denser the gratings, the higher the spatial frequency (white squares farther to the center in the frequency domain). Gratings with higher spatial frequency are more difficult to identify from a fixed viewing distance.
    }
    \vspace{-2em}
    \label{fig:bg_freq}
\end{figure}

\subsection{Image Frequency}\label{background:sp}

A two-dimensional (2D) array of pixels represents a digital image (the spatial domain), and each pixel corresponds to a particular color value (e.g., RGB). The representation of 2D pixels can be transformed into a sum of different frequency components in the frequency domain~\cite{bracewell2004fourier}. With the Fourier transformation, we can convert a digital image from a 2D spatial domain to a 2D frequency domain \cite{bracewell1986fourier} as shown in Figure~\ref{fig:bg_freq}. The frequency domain can be used to present the frequency distribution of an image. Additionally, the frequency of the image is determined by how fast the image pixel value changes. For example, in a fixed size figure shown in Figure~\ref{fig:bg_freq}, when there are more  black and white bars alternatively occurring in the gratings, the gratings' black and white pixel value changes more frequently, so the frequency of the gratings becomes higher. As image frequency increases, the human vision system 
(abbreviated as HVS) is harder to respond to the frequency. Thus, it is difficult for humans to distinguish between the alternative black and white bars when the number of them increases at a fixed image size (Figure~\ref{fig:bg_freq}).



\begin{figure}[ht!]
    \centering
    \includegraphics[width=\linewidth]{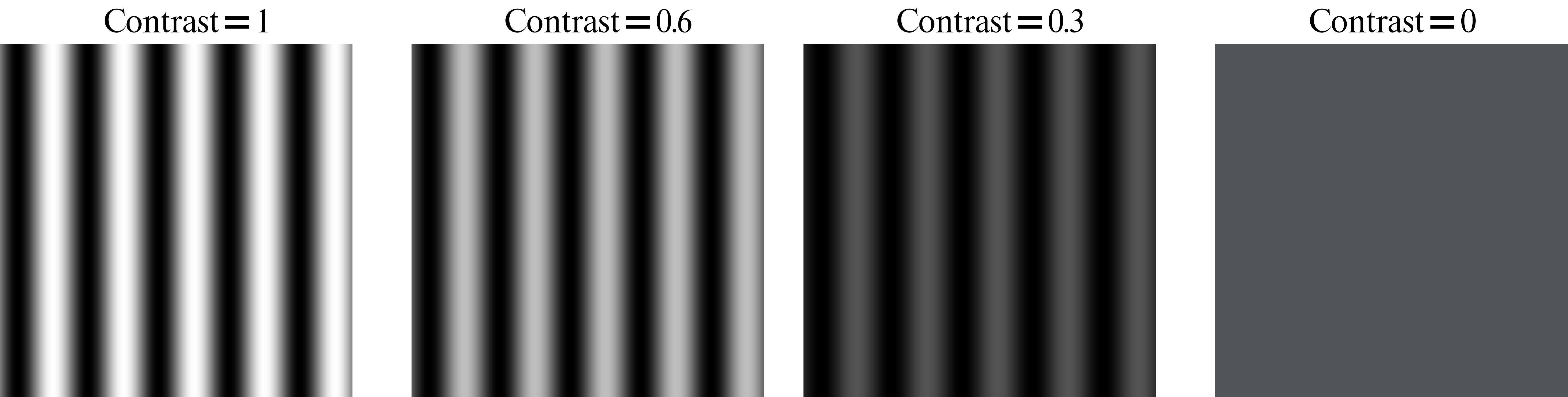}
    \caption{The effect of luminance contrast on human perception. Lower luminance contrast at a fixed viewing distance leads to poor visibility of the black and white gratings. }
    \vspace{-2em}
    \label{fig:bg_contrast}
\end{figure}

\subsection{Luminance Contrast}\label{background:contrast}
Luminance contrast refers to the luminance difference between pixels in an image. The HVS is more sensitive to the luminance contrast between pixels than the absolute luminance value of pixels. High contrast between an object and its background will enable the human to distinguish the object from the background. If the contrast does not reach the HVS detection threshold, humans cannot identify the object~\cite{lubin1997human}. For example, as shown in Figure~\ref{fig:bg_contrast}, when the luminance contrast between the white and black bars in the gratings decreases, we hardly identify the black and white bars. When the contrast reaches zero, we cannot perceive any bar in the image.

\begin{figure}[ht!]
    \centering
    \includegraphics[width=0.6\linewidth]{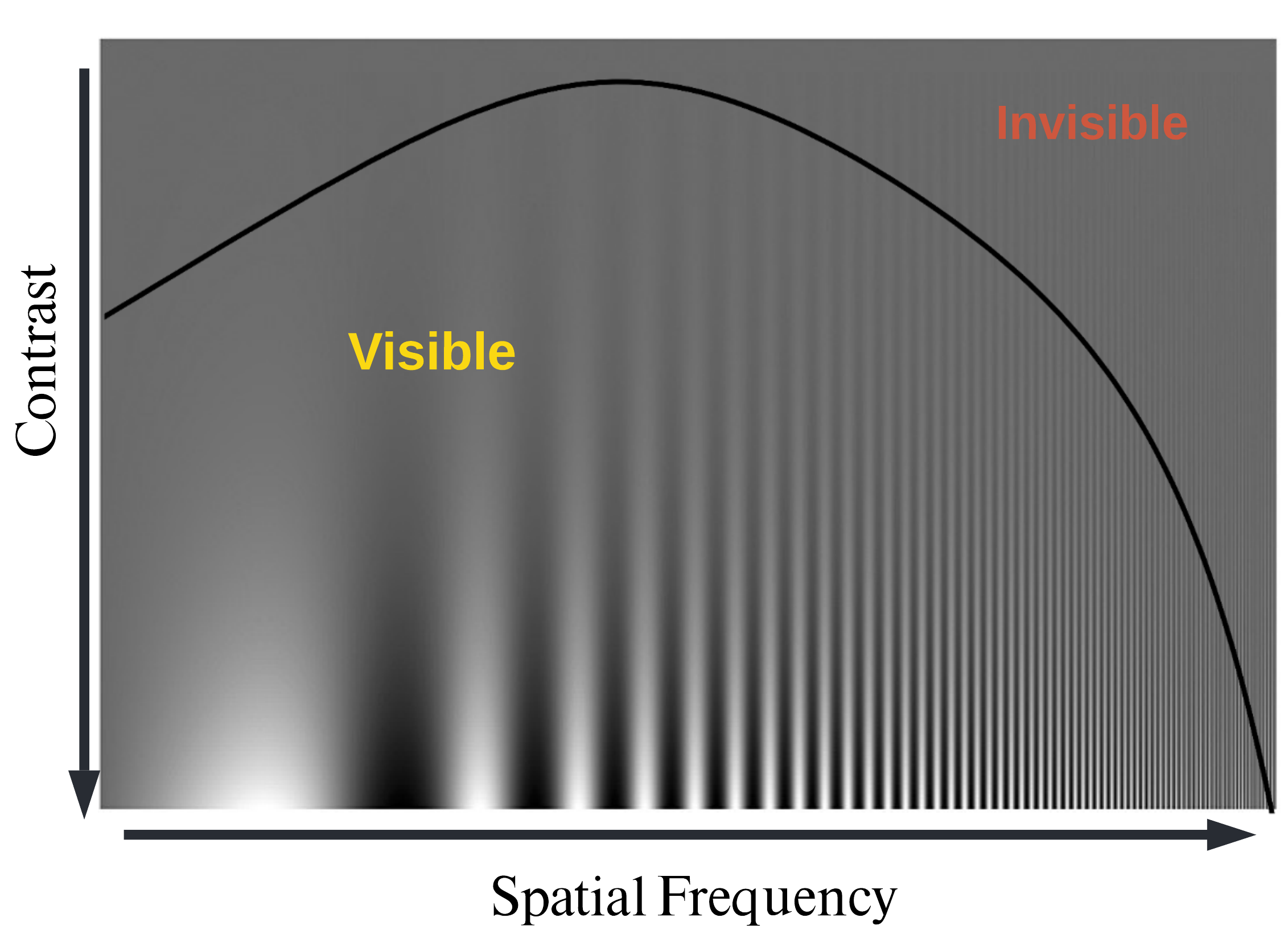}
    \caption{The contrast sensitivity function curve depicts the coupling effect between spatial frequency and luminance contrast on human visual perception.}
    \vspace{-2em}
    \label{fig:csf}
\end{figure}

\subsection{Contrast Sensitivity Function}\label{background:csf}

Though the HVS is affected by the contrast and frequency of the image, these factors do not influence human perception independently.
Viewing distance also affects human perception. To consider all these factors, researchers proposed a contrast sensitivity function (CSF).
Different from
image frequency, CSF utilizes the spatial frequency, which refers to the number of pairs of white and bar (Figure~\ref{fig:bg_freq}) on the retina at a given distance~\cite{sekuler1985perception}. For example, the image frequency can be regarded as the human view of the gratings at zero distance. When the human observes the gratings and moves away from them, the fixed-size gratings become smaller in the human eye, increasing the spatial frequency. In addition to the viewing distance, CSF assesses the HVS's contrast threshold over a range of spatial frequency~\cite{barten1999contrast}. Specifically, CSF indicates that the HVS exhibits different contrast thresholds at different spatial frequencies. As shown in Figure~\ref{fig:csf}, when the spatial frequency of the gratings becomes very high, the HVS requires a great contrast between bars. Otherwise, the human cannot identify those bars (the invisible area). Inspired by the CSF, our method leverages human vision characteristics to generate privacy-preserving visualization.

\begin{figure}[ht!]
    \includegraphics[width=\linewidth]{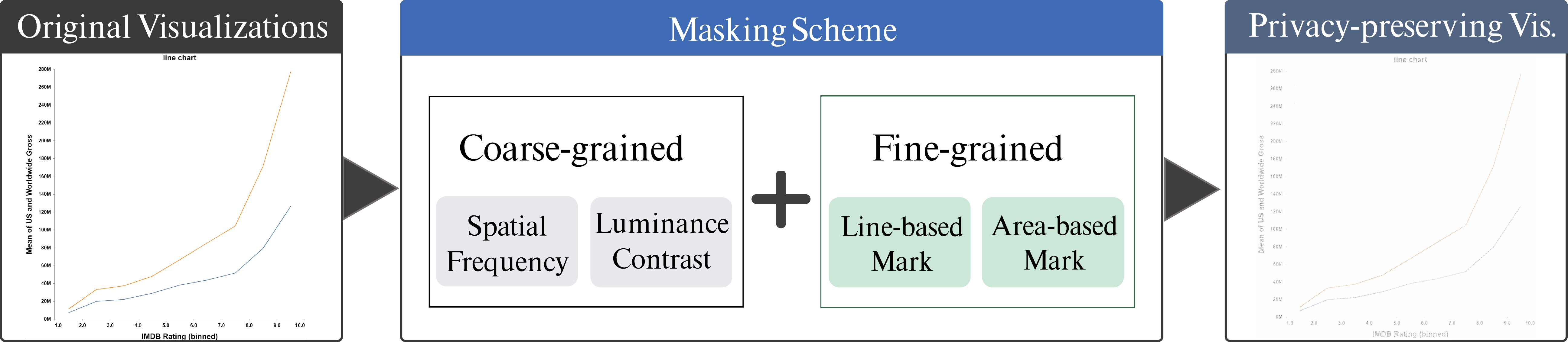}
    \caption{An overview of the proposed method. 
    It takes a visualization image as input and generates the privacy-preserving visualization.
    It comprises two levels of processing: coarse-grained masking and fine-grained masking.
    }
    \vspace{-2em}
    \label{fig:overiew}
\end{figure}

%% file: sources/4-method.tex
\section{Method}

Informed by prior research on human perception in Section~\ref{sec:background}, we propose a novel perception-driven approach to achieve privacy preservation for visualization on mobile devices.
Specifically, we present a masking scheme to process the bitmap image of an input visualization and transform it into a privacy-preserving one.
It consists of two major steps corresponding to two levels of processing granularity (Figure~\ref{fig:overiew}): \textit{coarse-grained masking} and \textit{fine-grained masking}.
Coarse-grained masking adjusts the spatial frequency of visual marks (Section~\ref{method:sp}) and their luminance contrast with the background in a visualization (Section~\ref{method:luminance}), which takes into account the fundamental principles of the human vision system.
Fine-grained masking further enhances the privacy preservation effect for visualizations by considering the distinct characteristics of different visual marks, which is informed by our Study 1 in Section~\ref{study:study1}.
Visual marks, such as circles in a scatter plot and bars in a bar chart, are fundamental elements in visualizations~\cite{munzner2014visualization,satyanarayan2015reactive,senay1994knowledge}. 
The source code for our approach is available online: \url{https://github.com/AlexanderZsh/Privacy-preserving-visualization}.

By considering the areas that different visual marks occupy,
we categorize visual marks into two types: \textit{line-based marks} such as lines, texts, and axes, and \textit{area-based marks} such as bars and circles. We propose adaptive fine-grained masking schemes for them to further improve the privacy preservation of visualizations (Section~\ref{method:improvement}).

\begin{figure}[ht!]
    \centering
    \includegraphics[width=1\linewidth]{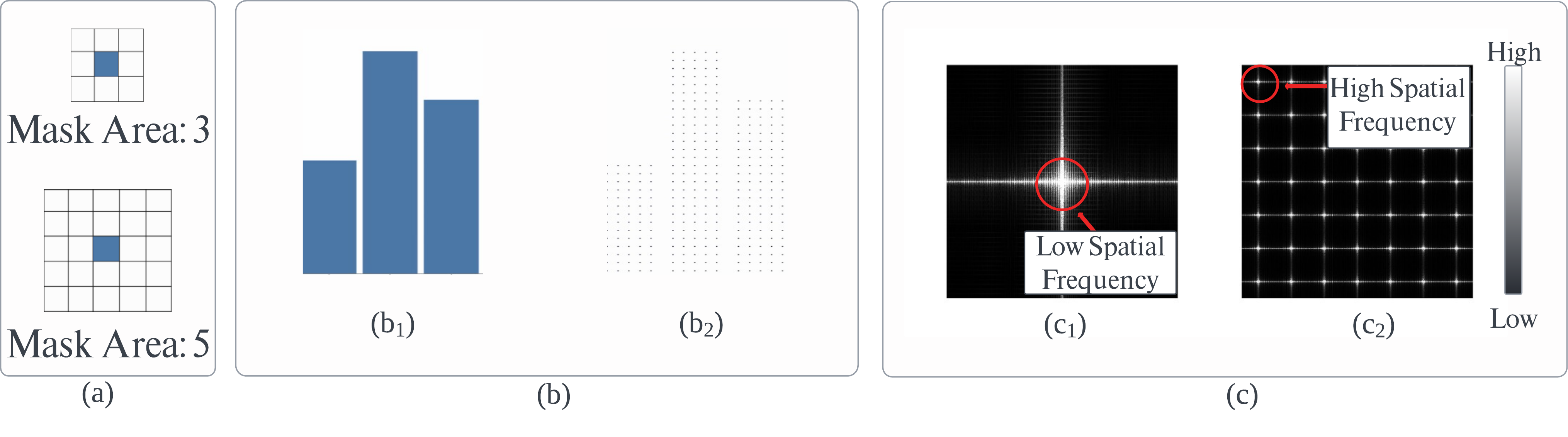}
    \caption{Area-based masking transforms area-based marks from low to high-frequency. (a) Examples of area-based masks with different mask areas. The pixel at the center of the mask remains unchanged, while the adjacent pixels are transformed to match the background color (e.g., white).
    (b) the effect of applying the area-based masking (Mask area: 5) to a bar chart ($b_1$) to convert it to its high spatial frequency version ($b_2$). (c) the frequency domain representations of ($b_1$) and ($b_2$).
    }
    \vspace{-2em}
    \label{fig:masking}
\end{figure}

\subsection{Coarse-grained Masking}

Coarse-grained masking aims to increase the spatial frequency and reduce the luminance contrast of visual marks in an input visualization to prevent shoulder surfers from viewing the visualization at a certain distance and allow visualization owners to see it clearly.

\subsubsection{Increasing Spatial Frequency}\label{method:sp}



Inspired by prior research on human vision (Section~\ref{background:csf}), we intend to increase the visual marks' spatial frequency using a binary mask.
First, we need to identify visual marks in an input visualization image. Given that visualizations usually have a white background, there is a clear color contrast between visual marks and the background of visualizations (Figure~\ref{fig:masking} (b$_1$)). In this paper, we leverage the Li Thresholding algorithm~\cite{li1998iterative}, which determines the color threshold between the background and visual marks. With the color threshold, we can identify visual marks from the background in visualization.
Then, we propose a masking scheme,
as shown in Figure~\ref{fig:masking}, to process which marks.
Such processing is called \textit{area-based masking} in this paper.
Specifically, we overlay the mask on the areas of visual marks, where the pixel at the center of the mask is retained, and other pixels are converted to the background color (e.g., white). The mask is tiled to cover the entire mark.
Take the bar chart in Figure~\ref{fig:masking} (b$_1$) as an example, the smooth bars will be converted to dotted bars with high spatial frequency (Figure~\ref{fig:masking} (b$_1$)).
The spatial frequency distributions of the bar chart before and after being processed by our masking scheme are shown in Figure~\ref{fig:masking} (c$_1$) and Figure~\ref{fig:masking} (c$_2$) respectively, indicating the increased spatial frequency of the processed visualization.
Accordingly, it makes it difficult for shoulder surfers to see the visualization at a distance, while users at a closer viewing distance can still clearly identify all the information 
from the processed visualization.
As the mask area increases, the bars
become more sparsely dotted, resulting in an increase in spatial frequency and making it harder for shoulder surfers at a distance to identify the processed bars.
\begin{figure}[ht!]
    \centering
    \includegraphics[width=0.8\linewidth]{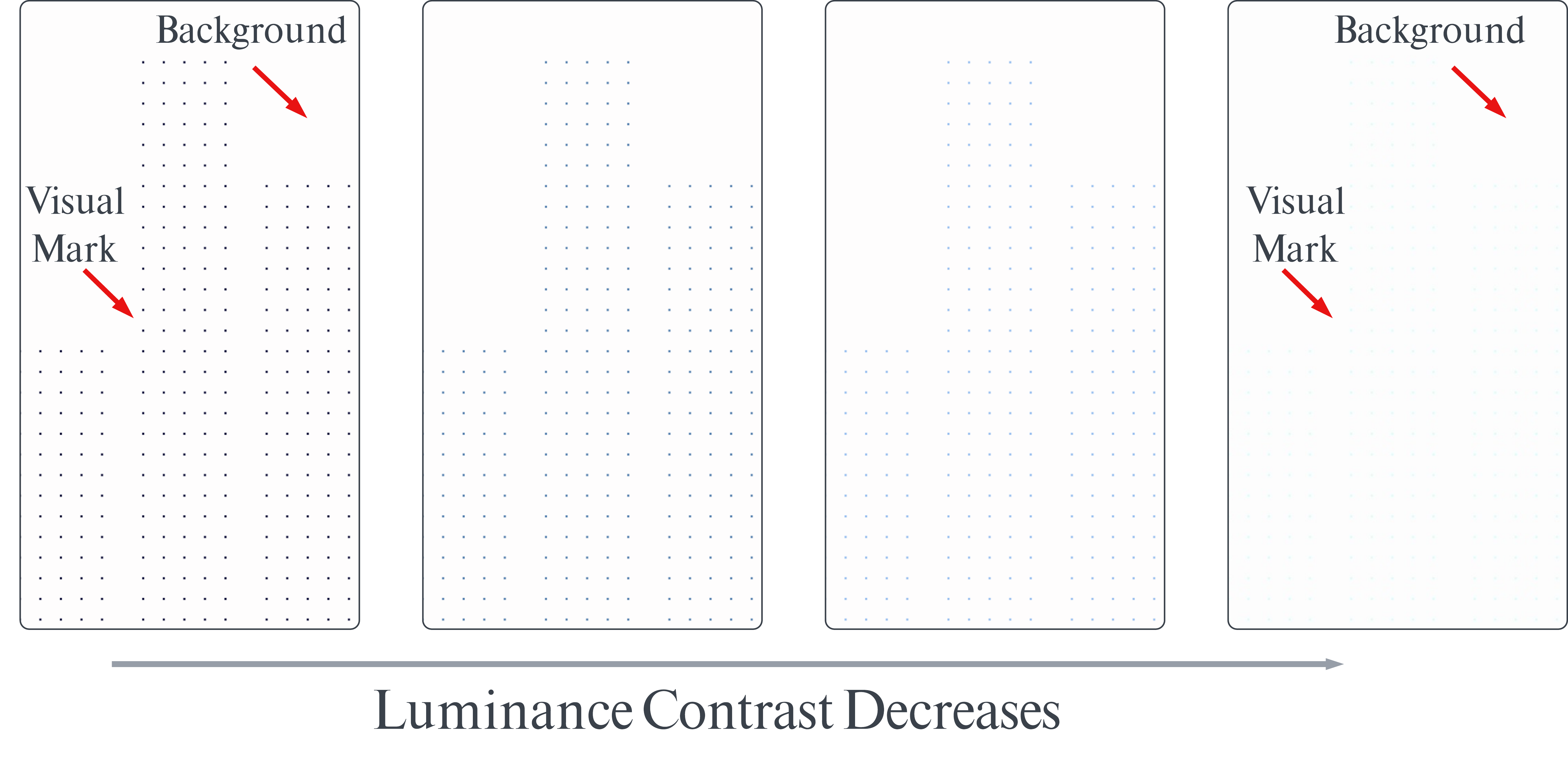}
    \caption{The effect of applying different luminance contrast on the converted high-frequency bar charts. The visibility reduces with the decrease in luminance contrast.
    When the luminance contrast reaches zero,
    we are not able to distinguish visual marks from the white background of a visualization.
    }
    \label{fig:luminance}
    \vspace{-2em}
    
\end{figure}

\subsubsection{Reducing Luminance Contrast}\label{method:luminance}

Besides adjusting the spatial frequency of visual marks, we also decrease their luminance contrast with the visualization background to further prevent shoulder surfers from seeing the visual marks of an input visualization.
Instead of using the commonly-used RGB color space, we employ the CIELAB color space when adjusting the luminance contrast of visualizations.
The major reason is that the RGB color space cannot accurately model how human perceives luminance~\cite{munzner2014visualization}, but the L channel of CIELAB color space aligns well with the actual perception of luminance by human vision system~\cite{hanbury2002mathematical}.
Therefore, by changing the luminance of the pixels of visual marks and background in the CIELAB color space, we can accurately control the luminance contrast between them that will be perceived by users.
Figure~\ref{fig:luminance} illustrates the influence of different luminance contrasts
in terms of CIELAB's L channel on the visibility of visual marks. 
With the decrease in the luminance contrast between the visual marks and the background, it becomes increasingly difficult for humans to distinguish visual marks from the background.

\subsection{Fine-grained Masking}\label{method:improvement}
\label{sec: further_improve_area}

The coarse-grained masking discussed above is designed to process all the visualizations without considering their own visual properties.
However, typical data visualization charts such as bar chart, pie chart, scatter plot, and line chart~\cite{battle2018beagle} consist of different visual marks and thus have distinct characteristics.
To further enhance privacy preservation performance, it is necessary for us to consider the unique visual properties of different visualizations.
Thus, building upon the coarse-grained masking, we further propose~\textit{fine-grained masking} to process input visualizations, as shown in Figure~\ref{fig:overiew}.
Specifically, we design adaptive masking schemes for line-based marks and area-based marks.


\begin{figure}[ht!]
    \centering
    \includegraphics[width=\linewidth]{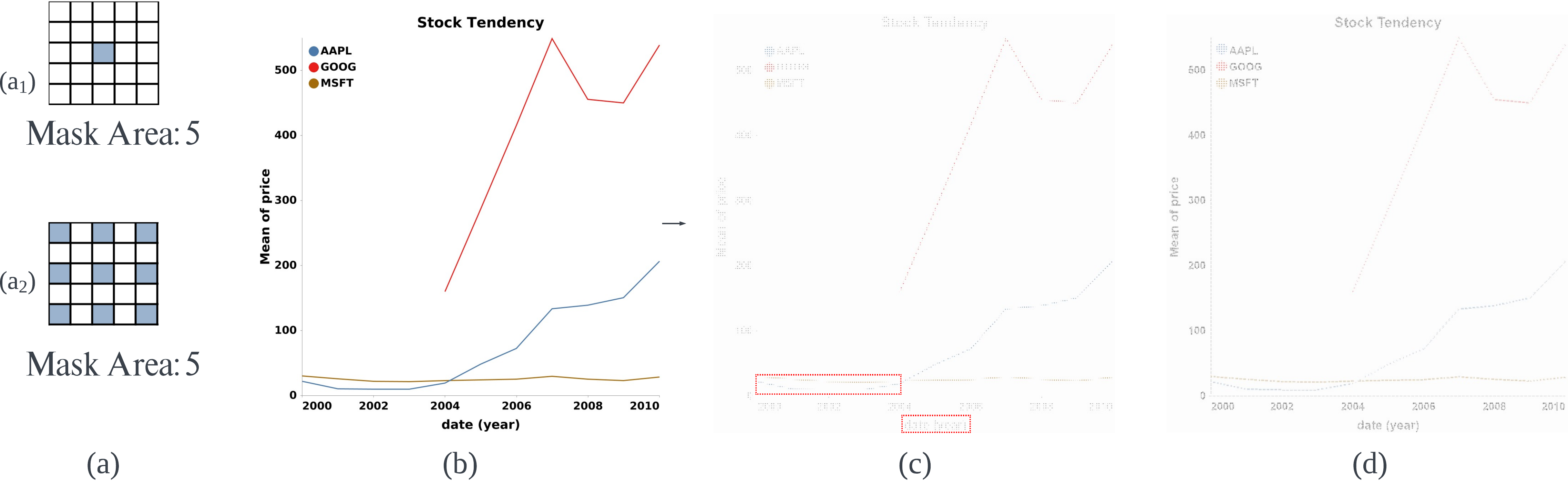}
    \caption{Line-based masking converts line-based marks from low to high frequency. 
    (a) An area-based mask with a mask area of 5 (a$_1$) and a line-based mask with a mask area of 5 (a$_2$).
    (b) an input line chart. (c) the line chart processed with area-based masking, where the dashed red rectangles highlight the problematic regions in the processing result.
    (d) the line chart processed with line-based masking, which retains more text and line information.
    }
    \vspace{-2em}
    \label{fig:line_masking}
\end{figure}


\subsubsection{Adaptive Masking for Line-based Marks}\label{sec-adaptive-mask-for-line}

As discussed above, we categorized visual marks into \textit{line-based marks} and \textit{area-based marks} due to their differences in the occupied areas.
During the development of our approach, we also notice that there is no one-size-fits-all solution that works well for both line-based marks and area-based marks of visualizations.
For example, Figure~\ref{fig:line_masking}(c) is the processed result of the input line chart (Figure~\ref{fig:line_masking}(b)) by using the area-based masking (Figure~\ref{fig:line_masking}(a$_1$)), where the masking scheme keeps only the pixel at the center. 
However, it is difficult to identify the lines and texts due to the obvious discontinuity in a few parts of these line-based marks (as shown within dashed rectangles in Figure~\ref{fig:line_masking}(c)). For some parts of the lines, the line segments are even broken, making it difficult to determine the trend of lines. 
Since the width of line-based marks (e.g., lines, axes, and texts) is often smaller than area-based marks, and it is essential to preserve the orientation of line-based marks,
we propose a new masking scheme for line-based marks, as shown in Figure~\ref{fig:line_masking}(a$_2$).
Figure~\ref{fig:line_masking}(d) shows the processed result of the input line chart by using the new masking scheme. Such processing is called \textit{line-based masking} in this paper.
By keeping more pixels surrounding the center,
it is clear to see that such a new masking scheme can better preserve the visual information of line-based marks while increasing their spatial frequency.

Furthermore, we adaptively adjust the size of the line-based masking for line-based marks according to their width. Among all the line-based marks (e.g., lines, axes, and texts), the width of lines and axes in data visualization charts are relatively stable and consistent. However, texts can vary a lot due to different font sizes, which motivates us to adaptively vary the size of the masking scheme to process texts specifically.
To this end, we first employ EasyOCR~\cite{jaided2020easyocr}, a widely used Optical Character Recognition
(OCR) tool to detect texts in the input visualization image.
Then, we further extract strokes of texts and determine text stroke width by 
using the fast parallel thinning algorithm~\cite{zhang1984fast} that has been integrated to the package Scipy~\cite{virtanen2020scipy}. 
The text stroke width is used to guide our empirical configuration on the adaptive mask size of our masking scheme for line-based marks.
 


\subsubsection{Customized Masking for Area-based Marks} \label{sec-mask4areamarks}


For the area-based marks like bars and circles, we initially leverage the area-based masking (Figure~\ref{fig:masking}(a)) to process them without specifically handling the borders of area-based visual marks.
As will be introduced in Section~\ref{study:study1}, we follow such a setup to evaluate 
the visibility of visualizations. The participants' feedback shows that the proposed area-based masking approach can achieve a good privacy preservation effect for area-based marks. However, it also makes it difficult for participants to accurately perceive the corresponding data of area-based visualizations due to the overly discretized borders of area-based marks. 
For example, for a processed pie chart, an excessively sparse border between two adjacent slices makes it difficult for human users to accurately identify the boundary between the two adjacent pie slices even at a close viewing distance, as shown in Figure~\ref{fig:study2_sample} (c) of Appendix~\ref{sec:appendix}.
To address this issue, we apply line-based masking,
as introduced in Section~\ref{sec-adaptive-mask-for-line}, to specifically process the borders of area-based marks, enhancing the accurate perception of area-based marks at a close viewing distance and guaranteeing privacy preservation for visualization above a certain viewing distance.

%% file: sources/5-experiment.tex
\section{Study 1: Variable Effects}
\label{study:study1}


By using the proposed masking scheme to process input visualizations, we can reduce the visibility of visual marks for shoulder surfers at a distance.
Meanwhile, we also need to guarantee that visualization owners nearby can still perceive the information from the visualizations. 
As observed from Figure~\ref{fig:csf}, the boundary is determined by both the spatial frequency and luminance contrast. 

Thus, we conducted Study 1
to investigate how these two factors jointly affect human perception of visual marks at different distances.
There are three primary objectives for Study 1:

\begin{itemize}
  \item Identifying how different mask areas affect human perception of the resulting visual marks.
  \item Examining how different luminance contrasts between visual marks and background affect human perception of visual marks.
  \item Studying how different viewing distances affect visual marks readability.
\end{itemize}


\begin{figure}[ht!]
    \centering
    \includegraphics[width=0.6\linewidth]{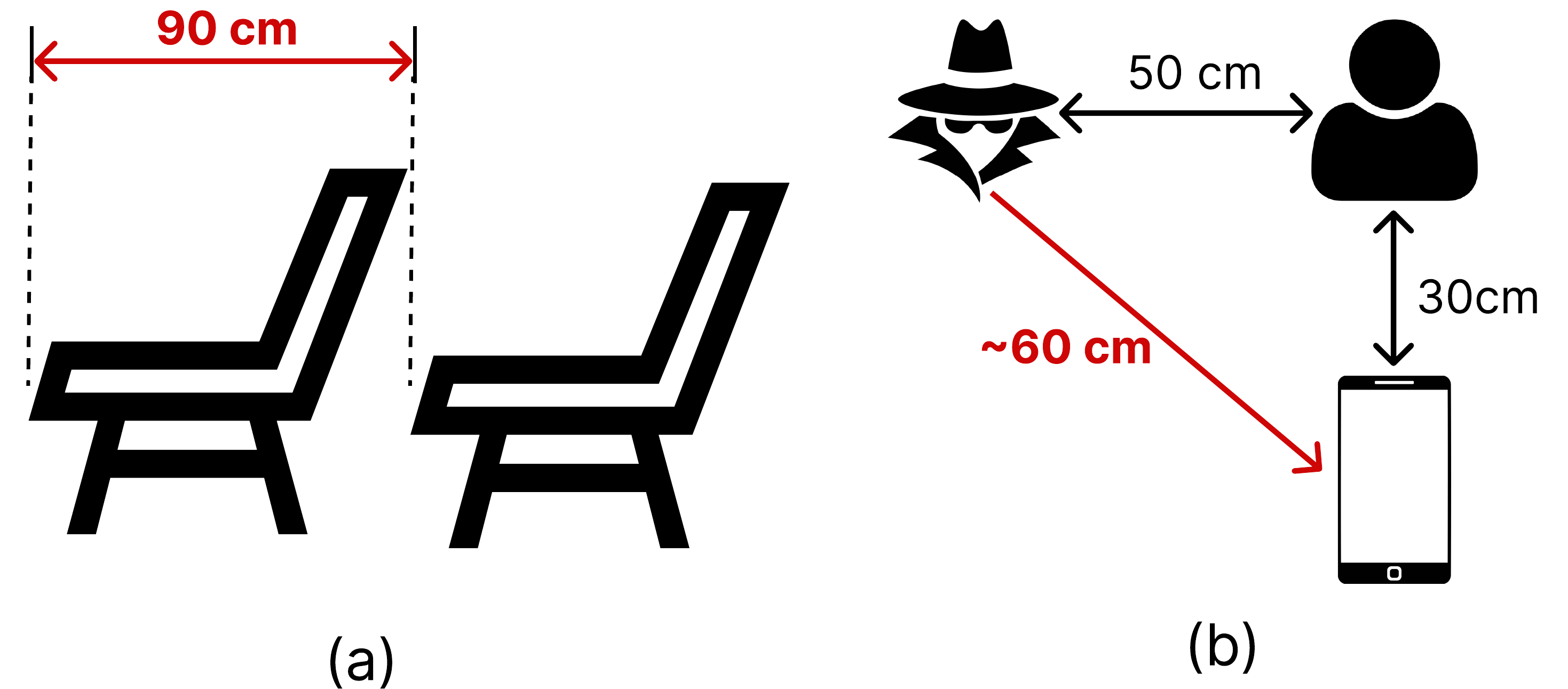}
    \caption{The viewing distance in various scenarios: (a) seat intervals, (b) shoulder surfer's distance from the user's phone while sitting side-by-side. 30cm denotes the distance between a mobile phone and users; 50cm represents average shoulder width; 60cm is the distance between a shoulder surfer and the user's phone when seated together; 90cm refers to the distance where the shoulder surfer sits behind the user in public transportation (e.g., bus).}
    \vspace{-2em}
    \label{fig:view_distance}
\end{figure}

\subsection{Participants and Apparatus}
All the participants are students recruited from a local university.
In total, 16 participants with normal or corrected-to-normal vision (3 females, age: 24-28) 
participated in the study. 
None of them was color blind.
The study was conducted in a campus meeting room with good lighting conditions.
Participants were requested to view a certain number of visualizations, and a short break was taken every 5 minutes to avoid visual fatigue. The whole process took around 40 minutes on average. Each participant was compensated with \$11. Before starting the study, participants read and signed an IRB-approved consent form.


 A 6.67-inches mobile phone with 1080 x 2400 pixels was placed on an adjustable stand on a table to accommodate different participants' heights. The authors switched the visualizations displayed on the mobile phone, and the participants sitting on a chair gave audio feedback indicating the visibility of each visualization.


\subsection{Stimulus}

As the bar chart, pie chart, scatter plot, and line chart are considered the most popular visualization types~\cite{battle2018beagle}, we evaluate the mask area and luminance contrast effect on all of them. For each visualization type, we consider seven mask area values for area-based marks (1, 3, 5, 7, 9, 11, 13) and line-based marks (1, 5, 9, 13, 17, 21, 25), respectively. In addition, we selected 5 luminance contrasts (0, 25, 50, 75, 100). To assure consistent visualization settings (e.g., test visualization size), we generated these testing visualization images
using \href{https://vega.github.io/editor/}{Vega-Lite}. Moreover, since the text is line-based by nature, we remove the text so that participants can concentrate on the marks. As a result, the test set consists of 140 visualizations (4 visualization types  $\times$ 7 mask areas  $\times$ 5 luminance contrasts). Some visualization examples used in the study
are shown in
Figure~\ref{fig:study1a_eg} in Appendix~\ref{sec:appendix}.

According to the prior study~\cite{yoshimura2017smartphone},
the viewing distance between users and the mobile phone screen is about 30cm. The length of the forward-facing seat is approximately 90cm (Figure~\ref{fig:view_distance}(a)), which is normal in public transportation (e.g., bus)~\cite{SeatWidth2021}.
Thus, we
set the distance between the participants and the smartphone with two values: 30cm (the viewing distance of the phone owner) and 90cm (the viewing distance of the shoulder surfer), as shown in Figure~\ref{fig:view_distance}(a, b).


\subsection{Procedures and Evaluation Criteria}
Since the visibility of visual marks
is the basis for visualization exploration,
we asked each participant
to view all visualizations 
one by one, and then rate the visibility of the marks from these visualizations. When reviewing relevant work~\cite{chen2019keep,papadopoulos2017illusionpin,lei2021iscreen}, we found that no prior work had developed
metrics to measure visibility.
Therefore, we proposed rating the visualization visibility with a 5-point Likert scale.
To ensure a more precise measurement of participants' subjective assessment of visualization visibility, our 5-point scale goes beyond simply indicating whether the visualization is visible or not. Instead,
we also consider the difficulty levels of identifying and perceiving the visualization in terms of the necessary effort and time when designing the criteria of the 5-point Likert scale. 
Specifically, the criteria for the 5-point scale are as follows: 

\begin{enumerate}
  \item[1:] I cannot recognize any visual marks from the visualization.
  \item[2:] I can identify a few visual marks from the visualization.
  \item[3:] I can identify a large portion of the marks from the visualization.
  \item[4:] I need some time and effort to identify all visualization marks from the visualizations.
  \item[5:] I can easily recognize all the visual marks at a glance.
\end{enumerate}

To calibrate participants' ratings, we provided participants with five different charts corresponding to the five scale scores before starting the real tests.
Participants were encouraged to explain why they rated the visualizations and what factors affect their visibility of the visualizations in a think-aloud manner. 
Given that we have set two viewing distances (30cm and 90cm), the viewing order might affect the rating. For instance, if the participant first viewed at 30cm, he/she is likely to see all the marks on the visualization, which may affect his/her evaluation of the visibility when viewing from 90cm.
To counterbalance this issue, we divided the participants into two groups, where the participants of Group 1 first viewed at 30cm and then 90cm, and vice versa for the participants of Group 2. Additionally, the visualizations are displayed to each participant in a random order to control order effects~\cite{strack1992order}.

\begin{figure*}[htp!]
    \centering
    \includegraphics[width=0.7\linewidth]{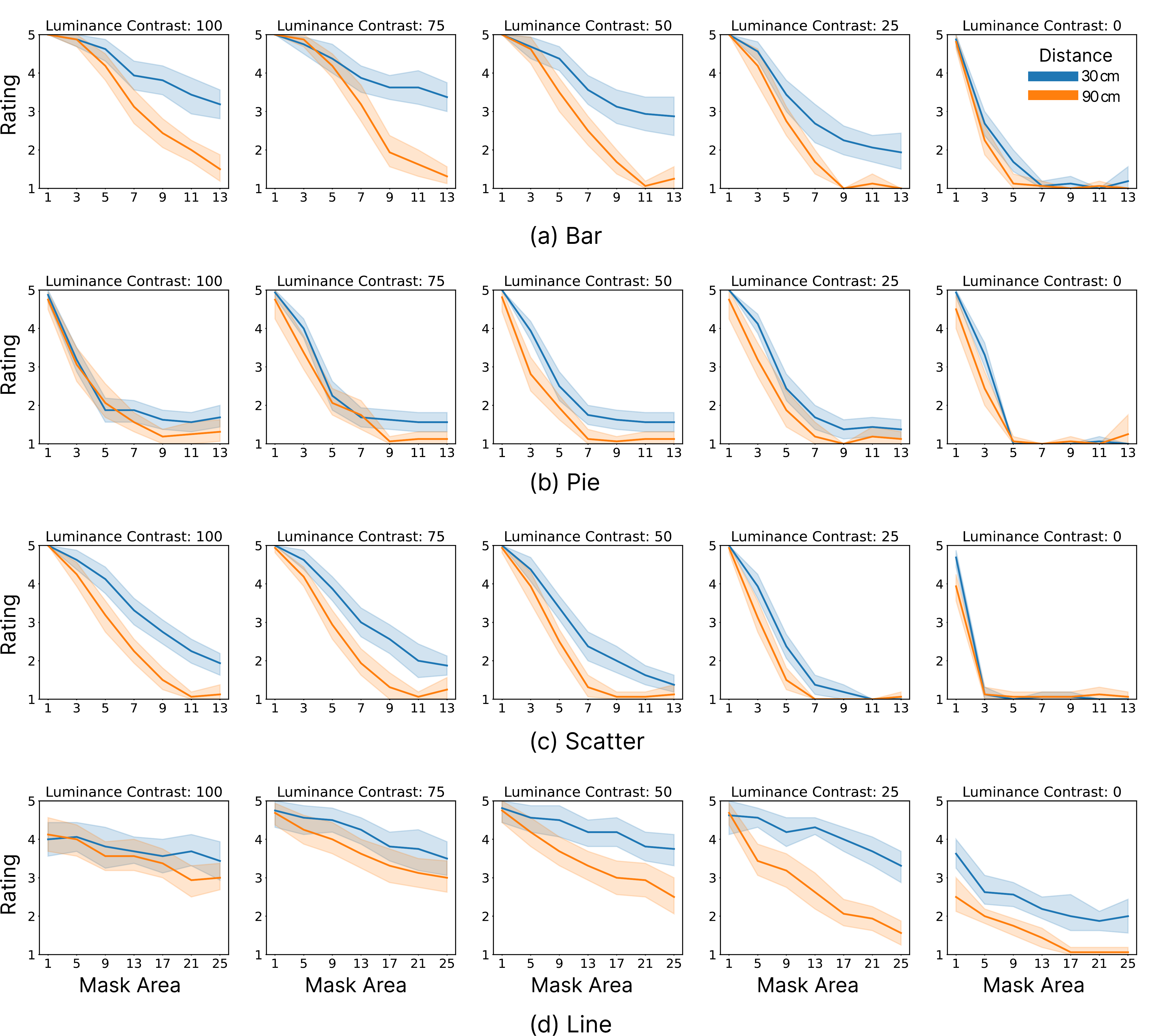}
    \caption{Average ratings obtained from Study 1, where each row corresponds to one visualization type. Different graphs in each row are for different luminance contrast. Each graph shows the effect of mask area size on visibility at close (30) and far (90cm) distances. The colorful bands (i.e., blue and orange) refer to the 95\% Confidence interval.}
    \vspace{-2em}
    \label{fig:study1a_rating}
\end{figure*}

\subsection{Result} Figure~\ref{fig:study1a_rating} shows the participants' average ratings under different settings (varying mask area and luminance contrast) for each type of visualization, from which we
have the following observations:

\textbf{Impact of Mask Area:} First, we can observe that participant ratings (i.e., visibility) decrease with the increase of mask area, regardless of the viewing distance, luminance contrast, as well as visualization type. The observation is reasonable as the increase of mask area actually enlarges the spatial frequency of the image, and human inherently shows poorer capability in seeing high-frequency content. Second, the decreasing rate of the visibility ratings for the two viewing distances is different in general. Specifically, viewing from a larger distance leads to a sharper decreasing rate, suggesting that spatial frequency has a more critical impact at a far distance.

\textbf{Impact of Luminance Contrast:}
Based on the CSF curve presented in Figure~\ref{fig:csf}, the visibility ratings are expected to decline with the decrease of the luminance contrast. However, as we can observe from Figure~\ref{fig:study1a_rating}, the ratings with a luminance contrast of 100 are consistently lower than those with a contrast of (75 or 50). The underlying reason is that: visualizations (e.g., bar charts) usually contain multiple visual marks with the same shape but with different colors (e.g., two bars correspond to different categories). With extremely low luminance, bars with different colors tend to exhibit the black color. As a result, although the visual marks are distinguished from the background, the information carried by different colors is lost. This implies that the CSF curve cannot be directly adopted in our visualization scenario, and we need to select the optimal luminance contrast based on the experiment carefully.

\textbf{Impact of Visualization Type:} When comparing the graphs (Figure~\ref{fig:study1a_rating}) in different rows, we can observe that although the overall trend of the curves is similar, the points where the curves start to converge are different. Thus, we have to select different variable values for different visualization types. Since our masking scheme aims to reduce the visibility at a far distance while maintaining the visibility in proximity, the optimal variable value corresponds to the case where the ratings between 30cm and 90cm have the most significant gap. Based on the result, the values of (mask area and luminance contrast) selected for bar, pie, scatter, and line chart are (13, 75), (7, 75), (5, 75), (21, 25).

\textbf{Indistinguishable Borders of Area-based Marks:}
Figure~\ref{fig:study1a_rating} shows that participants' visibility ratings for pie charts and scatter plots are quite similar across close and far distances.
By analyzing the feedback from participants, we find that the main reason is the indistinguishable borders of area-based marks in these visualizations.
Since the same area-based masking scheme (Figure~\ref{fig:study2_sample}(c) in Appendix \ref{sec:appendix}) is leveraged to process the whole area-based marks without specifically processing their borders, the borders of area-based marks are overly discretized, making it harder to distinguish the boundary of area-based marks.
For instance, P7 said \textit{"The dots [in the scatter plots] are small by nature. When a dot is divided into sparse pixels, it takes me more time to confirm whether these pixels belong to the same dot or not when viewing it at a close distance"}. P3 commented that \textit{"Even though I can recognize a pie chart [when viewing it at a close distance],
it was not easy for me to quickly identify each slice of the pie chart. Thus, I have lowered my visibility rating score at the close viewing distance."} 
Other participants have also reported similar issues with bar charts.
Such an observation is consistent with the Gestalt Principles~\cite{todorovic2008gestalt}, such as the Law of proximity and continuity.
It motivates us to propose customized masking for area-based marks, as introduced in Section~\ref{sec-mask4areamarks}. 



\section{Study 2: Privacy Preservation Effectiveness}\label{sec:study2}
Our second user study evaluated the effectiveness of our method in preserving privacy regarding mobile visualizations. This study aims to assess the complete version of our approach with suitable variables in comparison to the baseline approaches (i.e., original test visualization and our approach's partial version).


Following Study 1's recruitment process, we recruited 18 participants (5 females, Age: 22-28) with the normal or corrected-to-normal vision from a local university, ensuring none were color blind.
We maintained the same mobile device and room settings  and the IRB-approved consent form signing process. Each participant received a compensation of \$11 for their time in our study.

\subsection{Stimulus}
Similar to Study 1, bar charts, pie charts, scatter plots, and line charts were used in Study 2. Unlike Study 1, where we utilized one visualization chart for each visualization type, we used six charts for each visualization type to justify our method's effectiveness comprehensively. These chart images are generated by the Vega-Lite as well. In addition, we determined the variable values (i.e., mask area and luminance contrast) for each visualization type based on the results of Study 1. The variable values of mask area and luminance contrast are determined by its privacy preservation ability, namely, a high rating at a close viewing distance (30cm) and a low rating at a far viewing distance (90cm), as shown in Figure~\ref{fig:study1a_rating}. The variable values for each chart type are bar charts (mask area: 13, luminance contrast: 75), pie charts (mask area: 7, luminance contrast: 25), scatter plots (mask area: 5, luminance contrast: 75), and line charts (mask area: 21, luminance contrast: 25).


Aside from the two viewing distances tested in Study 1, we intended to further validate our method by adding a viewing distance that had not been tested. A common shoulder surfing scenario occurs when a user sits side-by-side with a shoulder surfer~\cite{says_fedup_2017}, so we add a different viewing distance (60cm) to account for such a scenario.
The reason why the viewing distance between the peeker and the phone is about 60cm is the average shoulder width is approximately 50cm~\cite{NASA} and the viewing distance of users is 30cm~\cite{yoshimura2017smartphone}, as shown in Figure~\ref{fig:view_distance}(a).

Our method aims to deter shoulder surfers from accessing visualization information on mobile devices. In this experiment, we used standard daily-use visualizations containing visual marks, axes, titles, and labels. There are three methods in the study:
\begin{itemize}
    \item \textbf{Original Visualization:} the original visualization is not processed by the privacy-preserving approach. 
    \item \textbf{Coarse-grained Visualization:} the visualization is generated by only the coarse-grained masking scheme in our method.
    \item \textbf{Fine-grained Visualization:} besides the coarse-grained masking scheme, the visualization is supplemented by the fine-grained masking scheme.

\end{itemize}

As a result, the test set consists of 60 visualizations (4 visualization
types × 5 charts × 3 methods). Some visualization examples used in the study
are shown
in Figure~\ref{fig:study2_sample} in Appendix~\ref{sec:appendix}.

\begin{figure}[ht!]
    \centering
    \includegraphics[width=\linewidth]{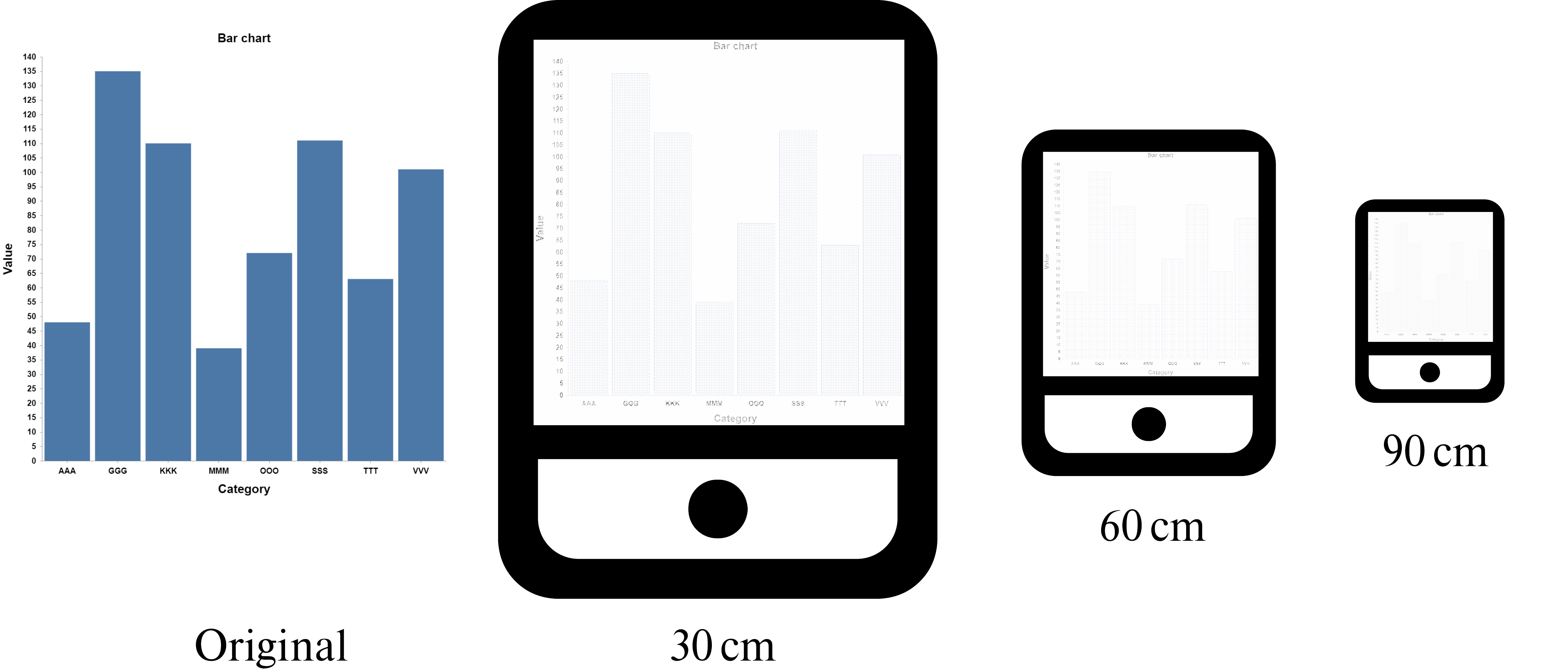}
        \caption{An illustration of our privacy preservation approach's effect on a visualization viewed at varying distances (30, 60, 90 cm), with the original bar chart shown on the left side.}
    \label{fig:final_effect}
\end{figure}

\begin{table}[!htp]
\centering
\begin{tabular}{p{1.5cm}|p{0.8cm}|p{5cm}}
\hline
\multirow{4}{*}{Motivation} &
\multicolumn{1}{l|}{Q1}  & Are you concerned about information leakage when using mobile visualization on a mobile device? If so, in what kind of condition? If not, why?        
\\ \cline{3-2}
 & 
\multicolumn{1}{l|}{Q2}  & Do you think our approach is helpful for privacy preservation while using mobile visualization on mobile devices?        
\\ \hline
\multirow{8}{*}{Effectiveness} &
\multicolumn{1}{l|}{Q3}  & From a close distance, are you able to see all the necessary information from our privacy-preserving visualizations? Why and why not?              \\ \cline{3-2} 
& 
\multicolumn{1}{l|}{Q4}  & From a long distance, are you able to see all the information? Why and why not?   
\\ \cline{3-2}
&
\multicolumn{1}{l|}{Q5}  & Can the border in the area-based visualization (i.e., pie, bar, scatter) can help you identify and receive the visual element?
\\ \cline{3-2}
&
\multicolumn{1}{l|}{Q6}  & Can you see the processed text in the close and far distance, respectively? Why and why not?
\\ \hline
Pros\& Cons &
\multicolumn{1}{l|}{Q7}  & What are the pros and cons of our overall approach?
\\ \hline
\end{tabular}
\caption{The questions in the post-study questionnaire of Study 2.}
\label{table:study2}
\end{table}

\begin{figure}[htp!]
    \centering
    \includegraphics[width=\linewidth]{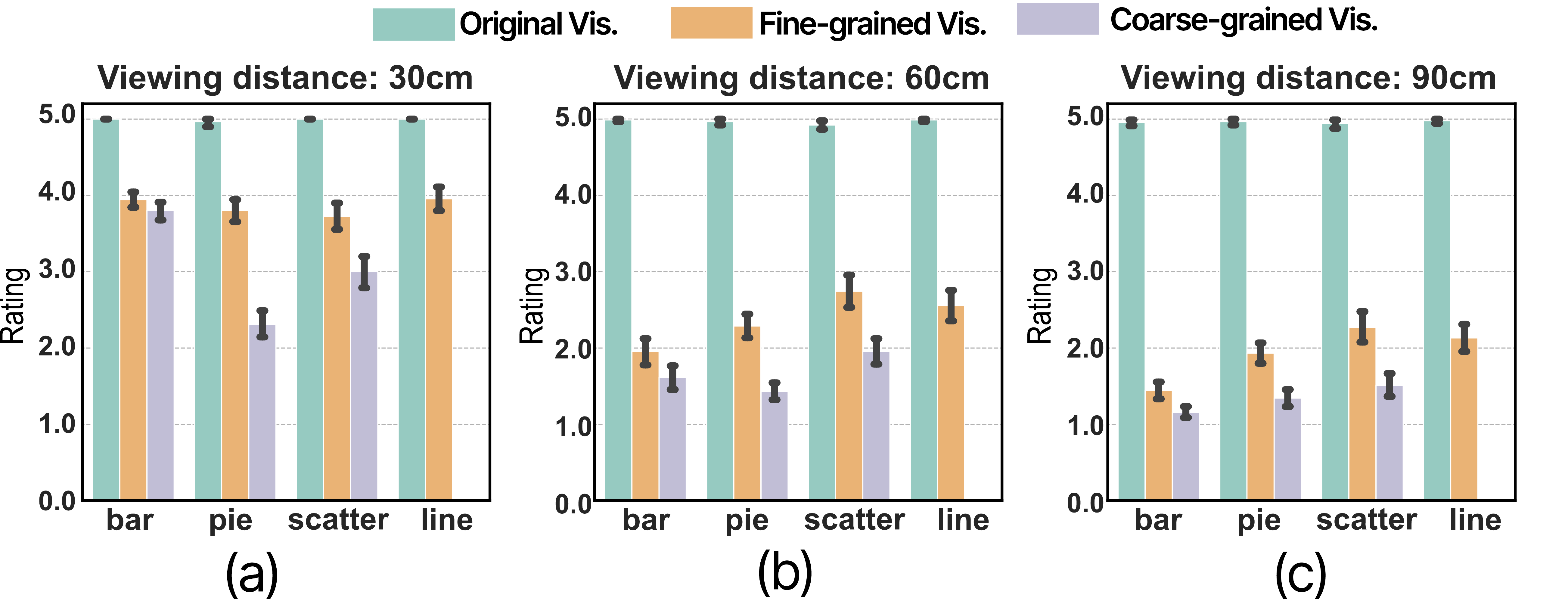}
    \caption{Study 2 average ratings for visual marks at various viewing distances: (a) 30cm, (b) 60cm, and (c) 90cm. The terms "Original Vis.", "Coarse-grained Vis.", and "Fine-grained Vis." represents the unaltered test visualization, visualization modified by coarse-grained masking, and test visualizations processed by both coarse- and fine-grained masking, respectively.}
    \vspace{-2em}
    \label{fig:study2_chart}
\end{figure}

\subsection{Procedures and Evaluation Criteria}
We compared our complete privacy-preserving visualizations with the other two baseline methods at three viewing distances. We conducted a within-subject study where tasks and methods were the factors.

The authors first introduced the five rating criteria to calibrate participants' ratings. The participants sat on a chair as the authors presented a series of visualizations
at three distances: 30cm, 60cm, and 90cm Figure~\ref{fig:final_effect}). Participants provided subjective evaluations for visual elements and text in the visualization at each distance on the 5-point scale. 
Furthermore, the displayed visualizations are in random order to control the order effects. 
According to the research by Poco~\textit{et al.}~\cite{poco2017reverse}, visualization consists of text (including axes) and visual marks. If the participants can clearly identify both text and marks, they are able to comprehend the visualization and vice versa. Therefore, there are two tasks for rating: visual mark visibility rating and text readability rating.

Participants entered their ratings into a mobile-based survey \href{https://www.qualtrics.com/}{tool}. The survey tool included a timer to record participants' time for each visualization rating. Participants were timed from when they started a visualization rating to when they finished it. The time is regarded as the participants' reading time to the visualization. A post-study interview was conducted to assess the quality of the study. The participants wrote their comments and suggestions in response to the open-ended questions in Table~\ref{table:study2} regarding our method's significance, usability, and overall feedback. All three distance combinations were conducted among participants in a counterbalanced manner. The study took about 60 minutes. The criteria for the 5-point scale are the same as study 1 (i.e., from 1-cannot see any data information to 5-see all data information clearly at a glance).

\subsection{Result}
Overall, participants can clearly see the visualization (i.e., visual marks and text) at a close viewing distance but hardly discern it at a far distance. Additionally, they need more time and effort to understand the visualization at a close distance. We will elaborate on the quantitative and qualitative results in the following.

\subsubsection{Quantitative Result}
\label{sec:study_2_results}
In visual mark visibility rating, Figure~\ref{fig:study2_chart} shows the participants' average rating for each of the three methods. Regardless of distances, participants can see the original visualizations. However, privacy-preserving visualizations (i.e., Fine-grained visualization) can prevent long-distance observing, as shown in Figure~\ref{fig:study2_chart} (b,c). For example, when participants are 60cm or 90cm away from the visualization, the average ratings of all visualization types are less than 3. According to the 5-point scale, it means that participants were able to view little visual marks on the privacy-preserving visualizations. In contrast, when the viewing distance reaches about 30cm, the average rating of our method is almost equal to 4, indicating that participants can see the visual marks without any difficulties. Furthermore, participants spent a minor amount of additional time (i.e., $\mu$=2.89s, $\sigma$=2.01s) getting information from (Coarse-grained + Fine-grained ) visualization compared to the reading time (i.e., $\mu$=0.82s, $\sigma$=1.07s) for original visualization. Also, Figure~\ref{fig:study2_chart}(a) demonstrates that the incorporation of borders can increase participants' visibility to the visual marks in close viewing distance, not decrease too much privacy-preservation of our method as shown in Figure~\ref{fig:study2_chart} (b,c).

\begin{figure}[ht!]
    \centering
    \includegraphics[width=\linewidth]{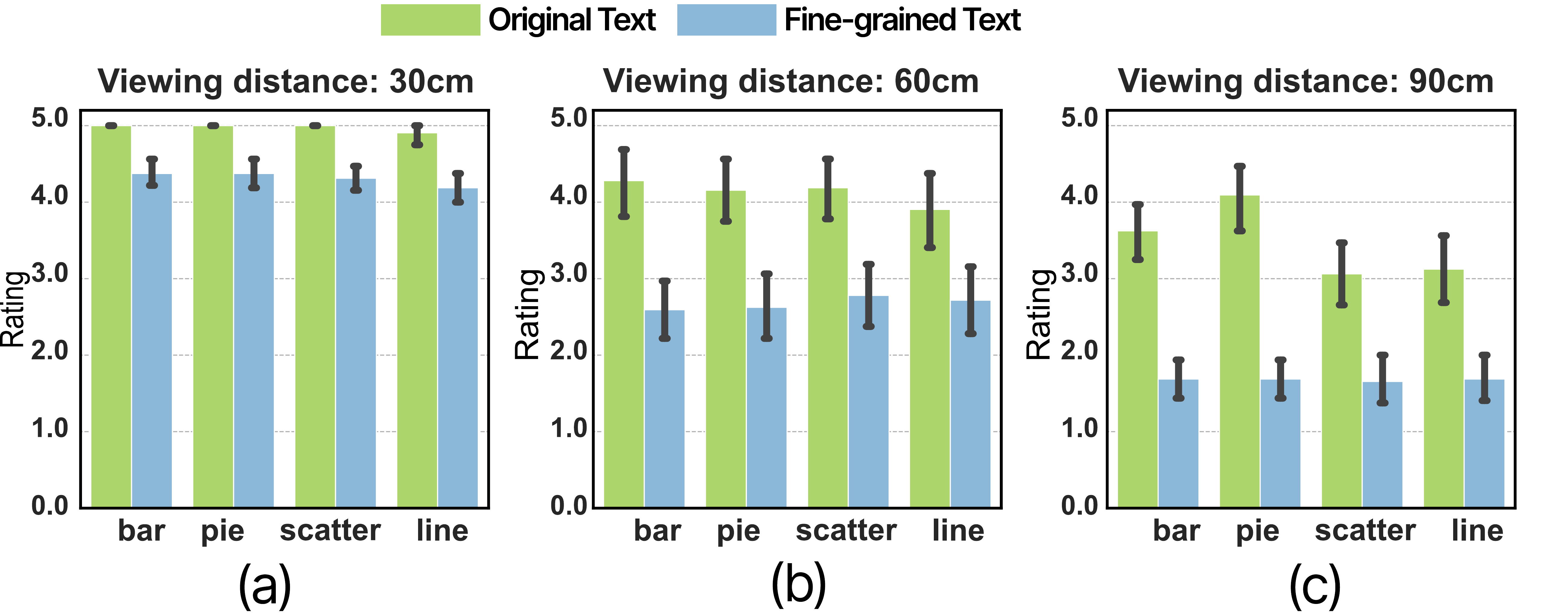}
    \caption{Study 2 average ratings for text at three viewing distances: (a) 30cm, (b) 60cm, and (c) 90cm. "Original Text" refers to the unaltered text in the test visualization, while "Fine-grained Text" corresponds to the text processed by both coarse- and fine-grained masking.}
    \vspace{-2em}
    \label{fig:study2_text}
\end{figure}

As for text readability, Figure~\ref{fig:study2_text} displays participants' ratings for the text in the original visualizations and processed text by our method. Figure~\ref{fig:study2_text} (a) verifies that participants can read the same content in the masked text as in the original text because the average rating significantly overcomes 4. As shown in Figure~\ref{fig:study2_text} (b,c), participants had difficulty identifying the text once the viewing distance increased, resulting in dramatic drops in their ratings. Similar to the chart, the reading time of the masked text (i.e., $\mu$=2.42s, $\sigma$=1.81s) is greater than the original text (i.e., $\mu$=1.04s, $\sigma$=1.05s), but the increase is acceptable in practice.

\subsubsection{Qualitative Feedback}

Overall, our privacy-preserving visualization received an enthusiastic response from the participants. Figures~\ref{fig:study2_chart} and~\ref{fig:study2_text} demonstrate the usefulness of privacy-preserving visualization. We summarized participants' feedback and categorized them into three groups in the following:

\textbf{The information leakage of mobile visualization is a common concern.}
Among all the eighteen participants,
fourteen of them expressed their concerns about the leakage of personal information shown as mobile data visualizations (e.g., it is not a good idea for anyone to view my screen regarding personal information). Seven participants (P7,11,13-16) expressed concern that the mobile visualization would disclose their financial information,

 \textbf{The proposed approach can guarantee privacy preservation after a distance.} 
All participants appreciated that privacy-preserving visualization helps protect their privacy on mobile visualizations. P11 could not distinguish the visualization from its background when viewing at a distance. P18, who had never viewed sensitive data on her phone due to a lack of trust in the phone's privacy protection, commented that \textit{"[Privacy-preserving visualizations] increase [my] trust in privacy being maintained (on the mobile devices)."} 

 \textbf{Area-based visualizations benefit from the added border.} Most participants recognized the usefulness of the addition border. P10 noted that when viewing at a distance, he could identify the visualization in detail (e.g., slices in a pie chart), but when he was far from the device, he only identified the visualization type. P16 emphasized the border effects on a pie chart, which enables him to determine each slice's size in the chart. P8 concluded that "the border can show the bounds of the information of data (encoded by the area-based marks)."


%% file: sources/6-discussion.tex
\section{Discussion}

The two
user studies
demonstrated the effectiveness of our proposed masking scheme on different types of visualizations. However, during the experiment, we also identified some limitations and lessons, which will be discussed in this section. 

\textbf{Trade-off between privacy-preservation and cognitive efforts.} As discussed in Section~\ref{sec:study_2_results}, there is a trade-off between privacy-preservation and cognitive efforts. Our proposed masking scheme successfully achieved privacy-preserving data visualization on smartphones, but it also requires slightly more time and effort for the data owners to read the visualizations processed by our approach.
Given that mobile data visualizations are often used to show personal data (e.g., bank account information and health data), which is often sensitive, we argue that better privacy preservation is much more critical compared with the slight increase of perception time and effort.

\textbf{Limited colors under low luminance.}
CIELAB color space represents a color using three dimensions, i.e., L, A, B, where L is the luminance~\cite{connolly1997study} and A, B represents the color hue. By adjusting the value of L, we can easily control the luminance contrast between the background and visual marks. However, since CIELAB color space is based on the human vision system, the luminance level and color hue are correlated~\cite{kim2009modeling}. More specifically, with the decrease in luminance L, the available color hue space is also reduced. Consequently, the change of luminance also modifies the color hue, resulting in a slightly different color~\cite{braun1999paradigm}. However, the users are still able to identify the visual marks~\cite{munzner2014visualization}.

\textbf{Parameter configurations for different devices.}
To control the number of variables in the user studies,
our evaluation of the proposed approach was conducted on a single mobile device and
the viewing angle of participants is fixed, i.e., participants were sitting right in front of the mobile device (at different distances).
However, similar to prior privacy-preservation methods~\cite{papadopoulos2017illusionpin,chen2019keep},
our privacy-preserving visualization approach also requires suitable parameter configurations (e.g., luminance contrast and mask areas) on different devices due to the varying screen sizes and resolutions.
We argue that it is a one-off effort, and the observations of our study results (especially Study 1) can guide the optimal configuration on new devices.
Also, it will be interesting to develop an approach to achieve automated optimal parameter configuration across mobile devices with varying screen sizes and resolutions and also take different viewing angles into account, which is left as future work.

\textbf{Generalizability.}
Our proposed method provides a flexible solution to enhance the privacy of mobile data visualizations without restricting the input visualization designs. It is capable of processing visualization images which is available for four commonly used visualizations and does not rely on any specific visualization generation tool/package.
Additionally, though our method is primarily designed for mobile visualization, it can be extended to visualizations on other non-mobile contexts with appropriate parameter settings, for example, visualizations on desktops used in public areas.

%% file: sources/7-conclusion.tex
\section{Conclusion}
In this work, we present a privacy-preserving approach to protect mobile data visualizations (e.g., financial and personal health data) from shoulder surfers in public spaces. Specifically, based on the characteristics of the human vision system, i.e., spatial frequency and luminance contrast jointly affect the human perception of an image when viewing from different ranges, we design a masking scheme that can be applied to process typical visualization charts like a bar chart, pie chart, scatter plot, and line chart. 
We conducted two
user studies
with 16 and 18 participants, respectively, to investigate the impact of different parameters, refine the devised masking scheme, and evaluate the real-world performance of our proposed method. The results validated the effectiveness of our approach in preventing information leakage from shoulder surfers.

In future work, we plan to extend our method to
more types of visualizations (e.g., customized visualizations) to further demonstrate its effectiveness in privacy preservation for data visualizations.
Also, it will be interesting to explore how we can leverage the screen setting of mobile devices (e.g., sizes and resolutions) to achieve automated optimal parameter configuration for our approach on different mobile devices.

%% file: sources/8-appendix.tex
\newpage
\setcounter{figure}{0}
\onecolumn
\appendix
\section{Test Visualization Examples}~\label{sec:appendix}
\begin{figure*}[h!]
    \hspace{-2em}
    \includegraphics[width=1\linewidth]{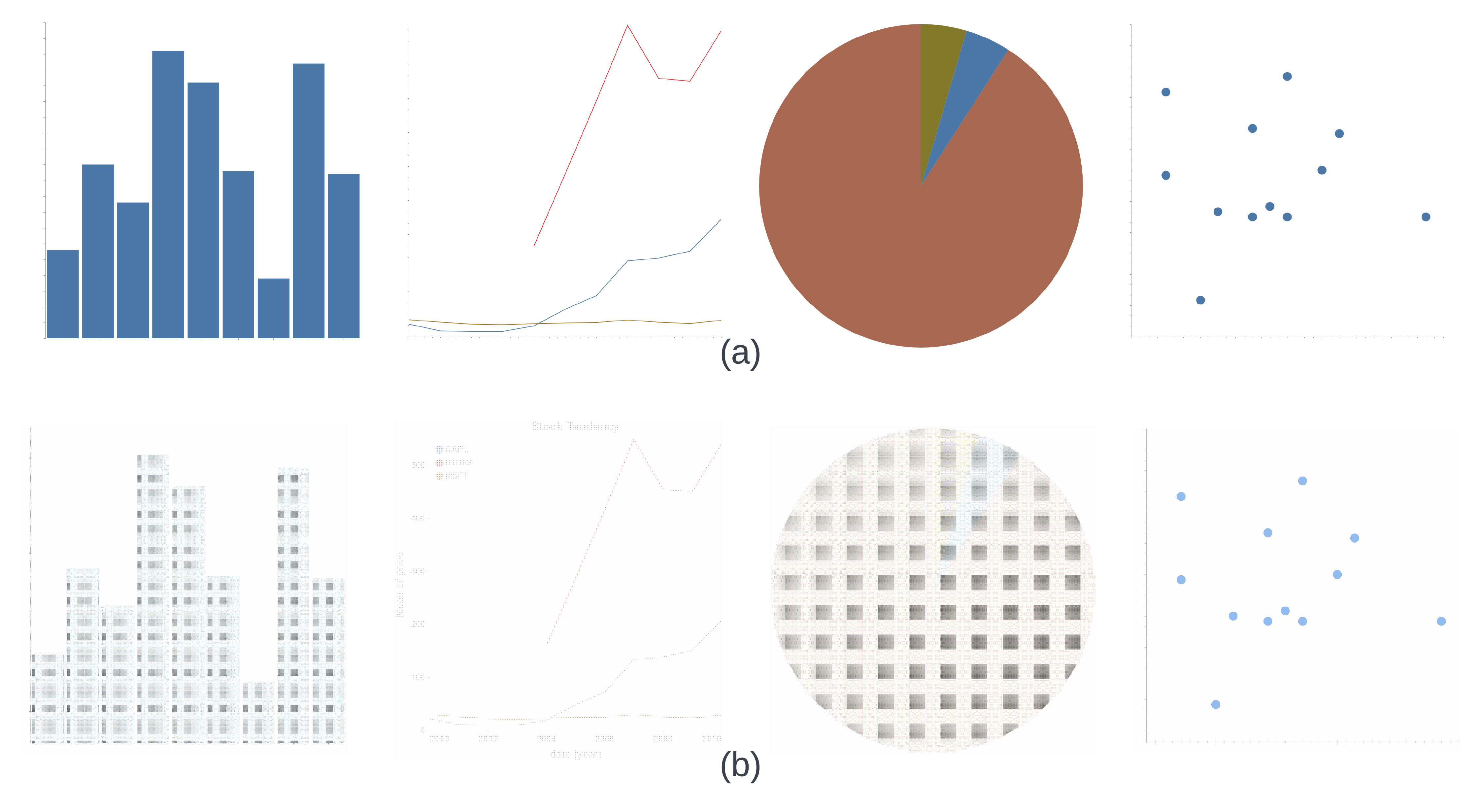}
    \caption{A comparison of (a) original visualizations and (b) privacy-preserving visualization processed by coarse-grained masking. Note: due to the varying screen resolutions, the displayed visualization effect may differ from that on the phones in Study 1.}

    \vspace{-1em}
    \label{fig:study1a_eg}
\end{figure*}

\begin{figure*}[ht!]
    \centering
    \includegraphics[width=\linewidth]{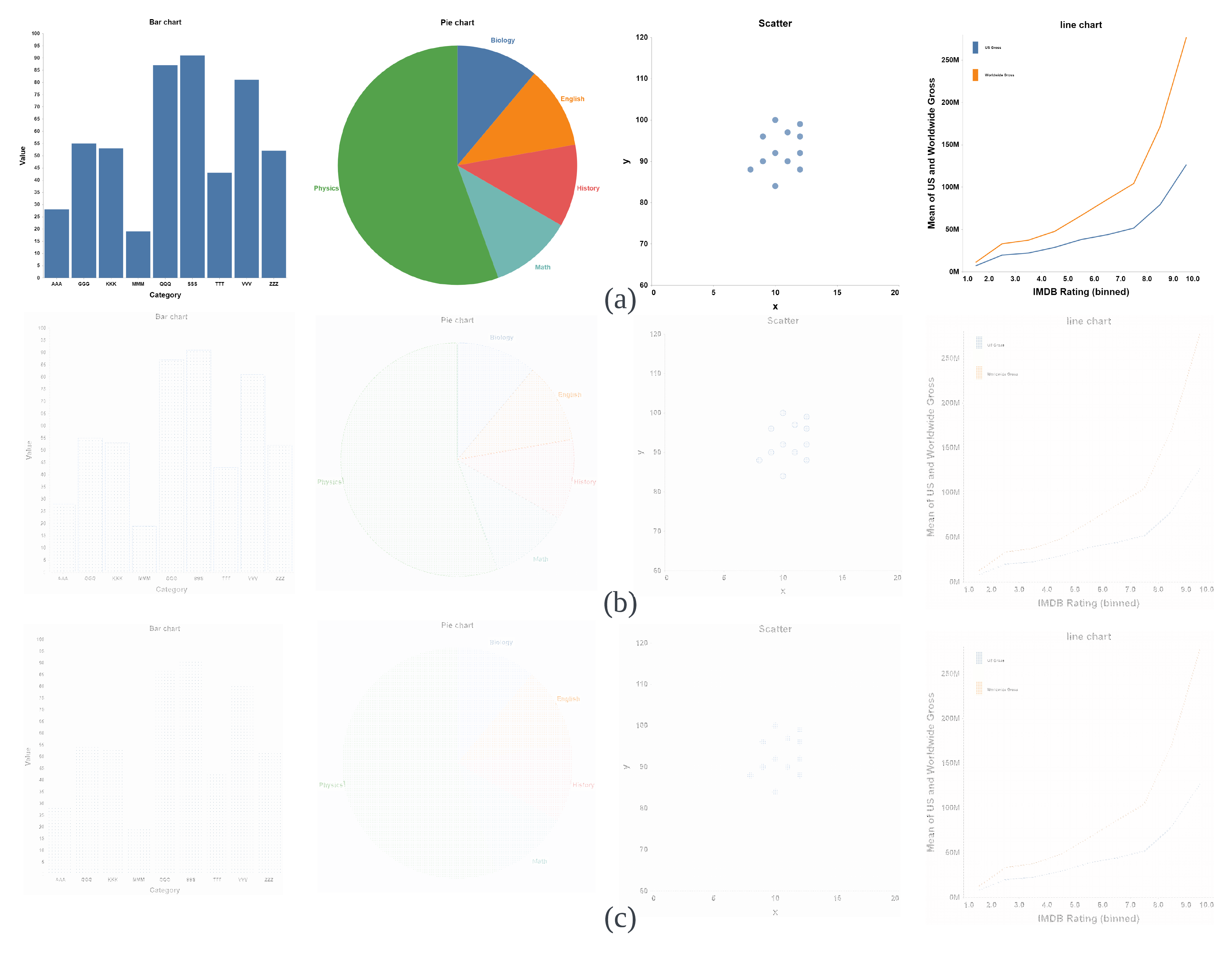}
    \caption{Four visualization samples employed in Study 2: (a) Original Visualizations, (b) Coarse-grained Visualization, produced by applying coarse-grained masking, (c) Fine-grained Visualization, generated through both coarse- and fine-grained masking techniques. Note: The visualization effect may vary due to different screen resolutions compared to those on the phones used in Study 2.}
    \vspace{-1em}
    \label{fig:study2_sample}
\end{figure*}

We display some visualization examples, including original and privacy-preserving visualizations used in our Study 1 (Figure~\ref{fig:study1a_eg}) and Study 2 (Figure~\ref{fig:study2_sample}). 

